**Electric-field control of oxygen vacancy and magnetic phase transition in cobaltite/manganite bilayer**

B. Cui,[1,2] C. Song,[1,*] F. Li,[1] X. Y. Zhong,[3] Z. C. Wang,[3,4] P. Werner,[2] Y. D. Gu,[1] H. Q. Wu,[5] J. J. Peng,[1] M. S. Saleem,[1] S. S. P. Parkin,[2] and F. Pan[1,*]

[1]Key Laboratory of Advanced Materials (MOE), School of Materials Science and Engineering, Tsinghua University, Beijing 100084, China

[2]Max Planck Institute for Microstructure Physics, Halle (Saale) D-06120, Germany

[3]Beijing National Center for Electron Microscopy, Laboratory of Advanced Materials and Department of Materials Science and Engineering, Tsinghua University, Beijing 100084, China

[4]Ernst Ruska-Centre for Microscopy and Spectroscopy with Electrons Research Centre Jülich, D-52425 Jülich, Germany

[5]Institute of Microelectronics, Tsinghua University, Beijing 100084, China

Manipulation of oxygen vacancies ($V_O$) in single oxide layers by varying the electric field can result in significant modulation of the ground state. However, in many oxide multilayers with strong application potentials, e.g. ferroelectric tunnel junctions and solid-oxide fuel cells, understanding $V_O$ behaviour in various layers under an applied electric field remains a challenge, owing to complex $V_O$ transport between different layers. By sweeping the external voltage, a reversible manipulation of $V_O$ and a corresponding fixed magnetic phase transition sequence in cobaltite/manganite ($SrCoO_{3-x}/La_{0.45}Sr_{0.55}MnO_{3-y}$) heterostructures are reported.

[*] songcheng@mail.tsinghua.edu.cn; panf@mail.tsinghua.edu.cn.

The magnetic phase transition sequence confirms that the priority of electric-field-induced $V_O$ formation/annihilation in the complex bilayer system is mainly determined by the $V_O$ formation energies and Gibbs free energy differences, which is supported by theoretical analysis. We not only realize a reversible manipulation of the magnetic phase transition in an oxide bilayer, but also provide insight into the electric field control of $V_O$ engineering in heterostructures.

## I. INTRODUCTION

Oxygen vacancies ($V_O$) are inevitable in multivalent transition metal oxides (TMOs) [1,2]. The functionality of TMO materials in information storage [3,4], ferroelectrics [5,6], and catalysis [7,8] is strongly dependent on the concentration and dynamics of $V_O$. Thus, the manipulation of $V_O$ and the corresponding physical properties are important issues in TMOs. It has been demonstrated that varying the electric field is an effective route for controlling the electrical and magnetic properties of various oxides based on the charge accumulation/depletion [9–14]. Recently, however, a polarized ionic liquid (IL) has been used to generate a much higher electric field that could even confine $V_O$ and realize electronic and crystal phase transitions in the entire volume of various single oxide layers like $VO_2$, $WO_3$, $YBa_2Cu_3O_7$, $SmNiO_3$, $SrCoO_3$, and manganites [15–21].

Compared with single oxide layer systems, oxide multilayer systems are more commonly employed in various electronics or energy devices, such as multiferroic/ferroelectric tunnel junctions and solid-oxide fuel cells [22,23]. In these systems, both the formation/annihilation of $V_O$ under an electric field determine their performance: the drift of $V_O$ in the electrode layers, caused by ferroelectric polarization switching, contributes to the resistance modulation in the tunnel junctions of (La,Sr)$MnO_3$/Pb(Zr,Ti)$O_3$/(La,Sr)$MnO_3$ and (La,Sr)$MnO_3$/$BaTiO_3$/Co [24,25]; while the enhancement of oxygen diffusion in both the cathode [e.g., (La,Sr)$MnO_3$ and (La,Sr)$CoO_3$] and electrolyte [e.g., $ZrO_2$:$Y_2O_3$ and $SrFeO_{2.5}$] under an electric field results in the reduction of the operating temperature in a solid-oxide fuel cell [23,26]. As a result of the $V_O$ transport between different oxide layers, the electric field control of $V_O$ in oxide multilayers is much more complex than that in the single layer, especially for the priority of $V_O$ formation/annihilation in the different layers.

In this study, we use an electric field applied through an IL to control the $V_O$ and magnetic phase transition in cobaltite/manganite ($SrCoO_{3-x}$/$La_{0.45}Sr_{0.55}MnO_{3-y}$) heterostructures with high ionic transport ability [23]. As model systems, $SrCoO_{3-x}$ and $La_{0.45}Sr_{0.55}MnO_{3-y}$ not only offer the advantages of a well-understood structural defect chemistry and relationship between the electronic and ionic properties but also show different $V_O$ formation energies as well as Gibbs free energies in different oxidized states [Figs. 1(a) and 1(b)] [8]. Thus, the competition of $V_O$ formation/annihilation in these two layers is expected to be reflected by the magnetic phase transition sequence of the heterostructure under an electric field. This is hoped to provide a greater understanding of the electric-field-controlled $V_O$ behaviour in various electronic devices and chemical processes involving oxide bilayers.

## II. METHODS

All of the samples were grown using pulsed laser deposition (PLD) from stoichiometric $La_{0.45}Sr_{0.55}MnO_3$ and $SrCoO_3$ target by applying a KrF excimer laser. During the growth, the substrate was held at 750 ℃ and in an oxygen background pressure of 100 mTorr for $La_{0.45}Sr_{0.55}MnO_3$, 870 ℃, 10 mTorr for $SrCoO_{2.5}$ and 850 ℃, 200 mTorr for $SrCoO_3$. The ionic liquid N,N-diethyl-N-(2-methoxyethyl)-N-methylammonium bis-(trifluoromethyl

sulfonyl)-imide (DEME-TFSI) was used to apply gate voltage. When a positive (negative) gate voltage ($V_G$) is applied on the IL and thin film, positive (negative) charged ions in IL move to the interface between IL and thin film, which causes the accumulation of electrons (holes) on the side of thin film and results in a electric double layer (EDL) at the interface. The thickness of EDL is estimated to be around 1 nm, hence the electric field at the interface between IL and thin film can reach $V_G \times 10$ MVcm$^{-1}$ [or on the order of 10 MVcm$^{-1}$] [27]. Then the electric field could attract the charged oxygen ions or vacancies in the thin film, inducing the oxidation or reduction of thin film [28]. Samples with large areas of $5 \times 2.5$ mm$^2$ are needed and almost the entire films are gated by IL before corresponding scanning transmission electron microscopy (STEM), X-ray diffraction (XRD), and X-ray absorption spectroscopy (XAS) measurements. The film was patterned into Hall-bar structure with gate electrode located in the vicinity of channel (transistor devices) by photo-lithography and wet etching to do the transport measurements. The sketches for thin film and transistor devices for ionic liquid gating are shown in Figs. 1 (c) and (d), respectively. More details about the device fabrication and the properties of IL used here are shown in Fig. S1 of [29].

The crystal structure of the heterostructure was characterized by XRD (Rigaku Smartlab, Cu-$K_\alpha$). The atomic resolution aberration-corrected STEM-HAADF (high angle annular dark field) image were taken by JEOL JEM-ARM200F. STEM-ABF (annular bright-field) were performed in TITAN 80-300 STEM and G2 80-200. The XAS measurements in total electron yield mode were done at Beamline BL08U1A at Shanghai Synchrotron Radiation Facility (SSRF) (at 300 K). The conductivity and magnetroresistance were determined in the physical property measurement system (PPMS) with a constant in-plane current of 10 μA from 380 to 10 K. Gate voltage was applied by Agilent 2901A. All the transport experiments were carried out after maintaining $V_G$ for 30 min without special instruction. First-principle calculations were carried out with the projector augmented wave implementation of Vienna *ab initio* simulation package (VASP).

## III. RESULTS

Before investigating the complex bilayer system, we show the electrical manipulation of electronic phases in a single layer of $SrCoO_{3-x}$ (SCO) and $La_{0.45}Sr_{0.55}MnO_{3-y}$ (LSMO) in Figs. 1(a) and 1(b), respectively. Positive and negative $V_G$ would accumulate and deplete $V_O$ in the oxide thin film, respectively [20,21,30,31]. Hence, a positive $V_G$ drives SCO to be antiferromagnetic/insulating (AFI) brownmillerite $SrCoO_{2.5}$ (BM-SCO) and LSMO to the low oxidization level (L-LSMO), while a negative $V_G$ favours the ferromagnetic/metallic (FM) perovskite $SrCoO_3$ (P-SCO) and the highly oxidized LSMO (H-LSMO), as marked by the arrows in Fig. 1. The M-LSMO approximately locates at a position with the optimal magnetic and electrical properties in the phase diagram. When M-LSMO is oxygenated to H-LSMO or reduced to L-LSMO, the conductivity and magnetism of the sample, which are coupled together due to the double-exchange mechanism, are suppressed [32].

We investigated the dependence of the crystal structure on gate voltages in a heterostructure with 15 nm BM-SCO on the top and 6 nm L-LSMO at the bottom (BM-SCO/L-LSMO) using XRD, as shown in Fig. 2(a). For this characterization, the whole area of the heterostructure was gated by the IL and the nonvolatile nature of the gating effect guaranteed the feasibility of the *ex-situ* experiments after removing gate voltages [15,16,20,21]. This method was also adopted in the other *ex-situ* experiments of STEM and XAS. Gate voltages were applied in the sequence as follows: ① $V_G = 0$, ② $V_G = -3.25$ V, ③ $V_G = -4.5$ V, ④ $V_G = -5.75$ V, ⑤ $V_G = +1.25$ V, ⑥ $V_G = +3.5$ V, and ⑦ $V_G = +4.75$ V. The BM-SCO film clearly exhibits a characteristic doubling of the *c*-axis lattice constant that originates from the alternate stacking of octahedral and tetrahedral sub-layers along the *c*-axis. As a negative $V_G$ extracts $V_O$ from the system and drives the states from ① to ④, the diffraction intensity for BM-SCO gradually reduces and that of P-SCO increases in states ③ and ④. In contrast from the gradual appearance of the diffraction peaks for P-SCO, those of the BM-SCO phase are clearly observed once $V_G$ become positive (state ⑤). A further increase of the positive $V_G$ results in an increase of the diffraction intensity for BM-SCO as the ordered $V_O$ returned to the lattice. It should be noted that the diffraction peak from LSMO is absent because the thickness of LSMO is only 6 nm and its lattice parameter is close to that of the STO substrate ($c_{LSMO}$ = 3.87 Å, $c_{STO}$ = 3.91 Å). The diffractions of BM-SCO (004) and (008) are respectively submerged in the (001) and (002) diffractions of the substrates owing to their closed lattice parameters ($1/4c_{BM\text{-}SCO}$ = 3.93 Å). The IL gating controlled phase transition, oxygen content variation and corresponding dynamic process could be directly demonstrated by the *in situ* TEM results, as shown in Figs. S2–S5 of [29].

The left columns of Figs. 2(b) and 2(c) respectively show the cross-sectional STEM-HAADF images of heterostructures in two states of the whole phase transition sequence: the lowest oxidized state ① and the highest oxidized state ④, reflecting the typical microscopic structure of the macroscopic thin films in XRD measurements. Here, the alternate stacking of fully oxygenated octahedral and oxygen-deficient tetrahedral sub-layers are observed directly, which is consistent with the brownmillerite structure of SCO at state ① [Fig. 2(b)]. The local structural changes arising from the oxygen deficiency were visualized by the Co-Co lateral atomic spacing modulation (Co-Co pairs) within the tetrahedral layers owing to the volume expansion near the oxygen-vacancy sites (highlighted by the coloured spheres) [33]. Such collective displacements of Co ions in the tetrahedral layers created well-ordered one-dimensional vacancy channels, which formed a zigzag network on the (010) plane of BM-SCO. Thus, alternating $CoO_2$ and CoO sublayers were observed in the ABF image of state ① [right column of Fig. 2(b)], in contrast to the $CoO_2$ layers without oxygen vacancies as revealed by the ABF image in the right column of Fig. 2(c). The sites of $V_O$ located at the space between the Co-Co pairs, as marked by the red circles. The well-ordered vacancy channels are interesting for the study of fast ion conductivity in solid-oxide fuel cells. When a negative $V_G$ is applied to the sample, the ordered $V_O$ are neutralized. The crystal structure of state ④ is a uniform perovskite structure with fully oxygenated octahedral layers, as shown in Fig. 2(c), which agree with the XRD results. The crystal structure characterizations reveal that the electric field generated by the IL is an effective way to manipulate the oxygen vacancies.

The electrical modulation of $V_O$ not only changes the crystal structure of the heterostructure but also results in the variation of the electronic structure. Here, the XAS at 300 K shown in Fig. 3 (raw data in Fig. S6 of [29]) is used to illustrate the competition of $V_O$ formation/annihilation between SCO and LSMO. Because the attenuation depth of XAS in the total electron yield (TEY) mode is around 6 nm, the samples used in this section are SCO (5 nm)/LSMO (6 nm) and the absolute values of the gate voltages are slightly modulated to obtain a similar effect in SCO (15 nm)/LSMO (6 nm) according to the corresponding XRD results (see Fig. S7 of [29] for more details). The uniformed gating effect along the depth of the thin film also confirms the comparability of samples with different thicknesses (Fig. S8 of [29] and a previous report [21]). The peak positions of Co- (left axis) and Mn-$L_3$ (right axis) are summarized in Fig. 3(a), which reflect the valence of these two elements where the higher the peak position, the higher the chemical valence [34,35].

Interestingly, although the BM-SCO is on top of the heterostructure, which is nearer to the IL, the LSMO underneath exhibits a more obvious oxidization tendency as proved by the sharp increase in the Mn valence at a small negative $V_G$ (state ②). This result suggests that $V_O$ in BM-SCO is relatively difficult to be filled and $Co^{3+}$ is more stable than $Co^{4+}$. With a further increase of the negative $V_G$, the oxidizations of both SCO and LSMO become remarkable according to the increased Mn and Co valence. When a positive $V_G$ is applied (state ⑤), in contrast to the subtle decrease in the Mn valence, a dramatic decrease of the Co valence is observed. Although a further increase of $V_G$ in the positive direction makes the Co valence gradually decrease, the slow rate suggests that the reduction of SCO from P-SCO to BM-SCO is almost complete in state ⑤. In contrast, the reduction of LSMO is only observed under a larger positive $V_G$. After the whole $V_G$ sequence, both of the valences of Co and Mn almost return to the pristine state. Such a circulation of the valence state strongly indicates the reversibility of the electrical manipulation.

The O-*K* edge XAS in Fig. 3(b) provides the oxygen stoichiometry information of the heterostructures. The O-*K* edge XAS signals in the TEY mode are mainly contributed by the SCO and LSMO layers, rather than the STO substrate, after consideration of the exponentially damped XAS intensity with an increased distance from the thin film surface [36]. At the O-*K* edge, the ratio of peak A/peak B is related to the Co 3*d*-O 2*p* hybridization. As we normalize the curves by setting the intensity of peak A to be one, the higher the peak B intensity, the more $V_O$ exist in the systems [37]. Interestingly, the intensities of peak B are rather low in states ③ and ④ compared with the other states, indicating that the P-SCO is stable only when the negative gate voltage is relatively large. While the abruptly increased peak B intensity in state ⑤ indicates that the $V_O$ are favoured in SCO even at a small positive $V_G$. The XAS of Co-, Mn-*L* and O-*K* edges collectively indicate the fact that the $V_O$ prefer to remain in the SCO layer rather than LSMO. In the whole gating process, we did not observe any notable change in the surface morphology, suggesting that the thin film was not irreversibly damaged during the measurements (Fig. S9 of [29]).

Subsequently, we used first-principles calculations to prove the formation/annihilation priority of $V_O$ in the heterostructure. In the calculation, a superlattice structure $[(La_{0.5}Sr_{0.5}MnO_3)_2/(SrCoO_3)_2]_n$ with a sequence of $MnO_2$/LaO/$MnO_2$/SrO/$MnO_2$/LaO/$MnO_2$/SrO/$CoO_2$/SrO/$CoO_2$/ SrO/$CoO_2$/SrO/$CoO_2$/SrO is used [Fig. 4(a)], and the in-plane lattice parameters were fixed as $a = b = 3.905$ Å (the same as STO substrates). An energy cutoff of 500 eV and appropriate k-point meshes were chosen so that the total ground-state energies were converged within 5 meV per formula unit. The oxygen ions are classified into 16 different cases according to their chemical situations and marked by the numbers 1–16. The energies of the heterostructures with $V_O$ at different positions are calculated and are summarized in Fig. 4(b). The energies for $V_O$ at positions 11 and 13 ($CoO_2$ sub-layer) are the lowest in all of the 16 cases, which was consistent with the microstructure of BM-SCO observed in Fig. 2(b). Overall, the heterostructure with $V_O$ in bulk $SrCoO_3$ (positions 10–14) is more stable than the case with $V_O$ at the $SrCoO_3/La_{0.5}Sr_{0.5}MnO_3$ interface (positions 8, 9, 15, and 16), while the oxygen ions in $La_{0.5}Sr_{0.5}MnO_3$ (positions 1–7) is relatively difficult to lose. Thus, when a positive gate voltage is applied to the heterostructure, $V_O$ are produced in $SrCoO_3$ owing to the lower energy. For the reverse process, a negative gate voltage prefers to inject oxygen ions into $La_{0.45}Sr_{0.55}MnO_{3-y}$, rather than $SrCoO_{3-x}$, which strongly supports the formation/annihilation priority of $V_O$ in the $SrCoO_{3-x}/La_{0.45}Sr_{0.55}MnO_{3-y}$ heterostructures.

The reversible electrical control of the phase transition in SCO/LSMO heterostructure is reflected by the transport properties in the transistor device as shown in Fig. 5(a). The effective channel of the device is 400 μm long and 100 μm wide. The phases of $SrCoO_{3-x}$ and $La_{0.45}Sr_{0.55}MnO_{3-y}$ are determined according to the combination of the various crystal and electronic structures under different $V_G$. The values of $x$ and $y$ in different states are estimated from the XAS peaks in Fig. 3(a). The transport measurements are used to investigate the magnetic and electrical properties of different states controlled by the electric field and then to clarify the phase compositions in the heterostructure. The BM-SCO (15 nm)/LSMO (6 nm) heterostructure is fabricated into a transistor device. The resistance measured here is the parallel resistances of SCO and LSMO layer as shown in the inset of Fig. 5(b). The gate voltage is swept in the range between –4.25 and +4 V with a step of 0.25 V at 300 K, as shown in Fig. 5(b), and the dependence of resistance on the temperature (*R*-*T*) for seven inflection points are measured and are shown in Fig. 5(c). The application of $V_G$ was 30 min to stabilize each state and resistance variation in this period of time is shown in Fig. S10 of [29], which somehow reflects the dynamics of corresponding phase transition and oxygen transport in ionic liquid gating. Compared with the *ex-situ* experiments in the thin film samples, gate voltages with different absolute values but same tendency were used in the transistor device to obtain corresponding similar gating effects. The seven inflection points in the device could approximately correspond to the seven states in the films used for XRD or XAS measurements, because the transport properties in the Hall bars and magnetic properties in the films show comparable results at these points or states (Fig. S11 of [29]). Nevertheless, the absolute values of $V_G$ in the transistor device and the thin film are somehow different. We determined the Curie temperatures ($T_C$) of these seven states from the *R*-*T* curves [the metal-insulator transition (MIT) temperature] and summarize them in Fig. 5(d). The coercivity ($H_C$) and exchange bias fields ($H_{EB}$) could be measured by the dependence of the channel resistance on the magnetic field at 10 K, as shown in Fig. 5(e) ([38,39] and raw data in Fig. S11 of [29]). All the values for the abovementioned data are summarized in Table 1.

With the application of $V_G$ = –2.75 V, the oxygen vacancies in the channel are neutralized and the resistance of BM-SCO/L-LSMO (① $V_G$ = 0 V) decreases gradually from ~$6.5\times10^4$ to ~$2.6\times10^4$ Ω and the $T_C$ increases from 283 to 320 K, which suggests the formation of BM-SCO/M-LSMO (②). A larger negative $V_G$ leads to a dramatic drop of the resistance to $6.8\times10^2$ Ω at –3.75 V (state ③), accompanied with a notably increased $H_C$ and decreased $H_{EB}$ at 10 K. This suggests that the AFI BM-SCO is changed to be FM P-SCO. A further increase of the negative $V_G$ would drive the phase transition from M-LSMO to H-LSMO, which is reflected by the rise of the resistance in the range between –3.75 V and –4.25 V (state ④). Nevertheless, the $T_C$ value could not be read from the *R-T* curves any more in states ③ and ④, because the signal of LSMO is masked by that of FM P-SCO whose $T_C$ is not coupled with a metal-insulator transition [8].

Gate voltage is then gradually swept back to the positive direction, which introduces more $V_O$ in the states of ⑤–⑦. A sharp enhancement in the resistance of two orders ($5.1\times10^5$ Ω) around +1 V (state ⑤) is caused by the phase transition from P-SCO to BM-SCO. The insulating BM-SCO releases the contribution of LSMO to the transport property: the reappearance of MIT in the *R-T* curve locates at around 230 K is in line with the performance of H-LSMO in the classic phase diagram [32]. It is also noteworthy that the exchange bias is still rather weak owing to the antiferromagnetic nature of H-LSMO at low temperature. A large positive $V_G$ would drive the phase transition of LSMO from H-LSMO to M-LSMO (state ⑥, $V_G$ = +3 V) and then L-LSMO (state ⑦, $V_G$ = +4 V). The corresponding $T_C$ and $H_{EB}$ of these states are 317 K, 65 Oe (state ⑥) and 289 K, 75 Oe (state ⑦), respectively. In the entire IL gating process, there are five different but readily transferable sates (states ① and ② are close to ⑦ and ⑥, respectively), and the ratio of the highest and lowest resistances is close to $10^3$. The results in Figs. 5(b) and 5(c) are obtained by two individual $V_G$ sweepings, which are consistent with the resistance values (see Table S1 of [29]) and indicate the reversibility and reproducibility of the IL gating experiment. The determination of the SCO and LSMO phases in the bilayer is supported by the transport and magnetic properties of single SCO and LSMO with different oxidation levels (Figs. S12 and S13 of [29]).

## IV. DISCUSSION

The formation/annihilation priority of $V_O$ and resulting magnetic phase transition sequence in the $SrCoO_{3-x}/La_{0.45}Sr_{0.55}MnO_{3-y}$ heterostructures could be understood by a combination of the $V_O$ formation energy ($E_V$) [40,41] and Gibbs free energy difference ($\Delta G$)[8]. The $E_V$, $\Delta G$, and the resultant phase transition barriers ($E_B$) in SCO and LSMO are shown in Fig. 6. As both LSMO and SCO exhibit a high ionic mobility [23], $V_O$ can be injected into or extracted from the whole heterostructure if $V_G$ is sufficiently high to overcome the $E_B$ for the phase transition in both the top and bottom layers. In this situation, the phase transition with a negative $\Delta G$ (reduced total energy of the whole system) is favoured independent of the distance that the oxide layer is from the IL. In contrast, if $V_G$ is only high enough to overcome the $E_B$ for the phase transition in the top layer, this phase transition will occur whatever the value of $\Delta G$. As the oxygen ions must go through the top layer to the bottom layer, the phase transition in the bottom layer could occur only after the $E_B$ of the phase transition of the top layer has been overcome.

In the present case, a negative $V_G$ induces the phase transition of BM-SCO → P-SCO and L-LSMO → M-LSMO → H-LSMO. For BM-SCO/L-LSMO, once gate voltage overcomes the $E_B$ of 2.32 eV for the top layer, both the phase transitions of BM-SCO → P-SCO and L-LSMO → M-LSMO ($E_B$ = 2.18 eV) are possible. The L-LSMO → M-LSMO transition with a negative $\Delta G$ of –0.37 eV occurs prior to BM-SCO → P-SCO ($\Delta G$ = 0.62 eV) [8]. The M-LSMO → H-LSMO transition emerges only after the two transitions above because of the higher $E_B$ of 2.55 eV, which requires a larger $V_G$. Positive gate voltages drive the phase transition of P-SCO → BM-SCO and H-LSMO → M-LSMO → L-LSMO. As the $E_B$ of the surface P-SCO → BM-SCO is only 1.70 eV, such a phase transition is preferential to occur first, and then the phase transition of H-LSMO → M-LSMO → L-LSMO ($E_B$ = 2.70 eV) at the bottom could be activated at a higher gate voltage [40,41]. This qualitative model could be used to understand the sequence of the phase transition.

We designed a control experiment using a heterostructure with an inverse structure where 5 nm M-LSMO was grown on 15 nm P-SCO to further verify the role of the $V_O$ formation energy and Gibbs free energy difference on the IL gating effect. In this structure, the P-SCO was first grown to guarantee the high-quality epitaxy of LSMO and the pristine state was M-LSMO/P-SCO. The BM-SCO/L-LSMO and M-LSMO/P-SCO were cooled to room temperature in the oxygen of 1 Torr quickly and in 300 Torr at a rate of ~5 ℃/min, respectively, to get needed structures and compositions. Figure 7(a) illustrates the electric-field-controlled manipulation of the phase transition in LSMO/SCO according to the XAS peak shift in Fig. 7(b) (at 300 K and raw data in Fig. S14 of [29]). The most obvious feature is that the reduction of SCO from P-SCO to BM-SCO is faster than that of LSMO under positive $V_G$ even when the SCO is on the bottom of the heterostructure, as indicated by the states ②' and ⑦' in LSMO/SCO. Once the positive $V_G$ is sufficiently high to overcome the $E_B$ of the surface H-LSMO → M-LSMO → L-LSMO (2.7 eV or 2.55 eV), the P-SCO → BM-SCO transition ($E_B$ = 1.7 eV) is also possible and would occur first owing to its negative $\Delta G$ = –0.62 eV. In contrast, once the negative $V_G$ overcomes the $E_B$ in the LSMO top layer, the oxidation phase transition, which possesses a negative and positive $\Delta G$ in LSMO and SCO layers, respectively, is preferential in the LSMO. The reversible phase transition in this reversed structure under electric field is also proved by the $V_G$ dependent resistance (Fig. S15 of [29]).

Compared with the field effect transistors consisting of solid-dielectric gates such as $SiO_2$ and $SrTiO_3$, a much larger amount of oxygen vacancy formation/annihilation is found in polarized ionic liquid, where the large electric field (>10 $MVcm^{-1}$) [27] and possible redox reaction [28] at the ionic liquid/oxide interface might play crucial roles. Although the transport measurements were done with the application of gate voltage, the electrostatic effect is difficult to penetrate into such a thick heterostructure of SCO (15 nm)/LSMO (6 nm) due to the strong screening effect by the surface charge. For instance, the penetration thickness for the electrostatic effect is confined to a limited depth of atomic dimensions for metallic materials [4]. Moreover, the results of XRD, (S)TEM, and XAS experiments were carried out after the removal of gate voltages, excluding the contribution from electrostatic effect in ionic liquid gating. We believe that such a $V_O$ formation energy and Gibbs free energy difference controlled phase transition shows the potential to be generalized for other oxide bilayers with good ionic transport abilities. However, it should also be mentioned that for some systems with low ionic transport ability, like ferroelectric materials, the situation might be different (e.g., BTO/LSMO and PZT/LSMO) [34,42].

## V. CONCLUSION

In conclusion, the reversible manipulation of the magnetic phase transition in SCO/LSMO heterostructure is realized based on the oxygen vacancies formation and annihilation under an electric field. In such a phase transition, the brownmillerite $SrCoO_{2.5}$, perovskite $SrCoO_3$, high-, middle-, and low-oxidization-level $La_{0.45}Sr_{0.55}MnO_{3-y}$ (H-LSMO: $La_{0.45}Sr_{0.55}MnO_{2.99\sim2.96}$, M-LSMO: $La_{0.45}Sr_{0.55}MnO_{2.91\sim2.85}$, L-LSMO: $La_{0.45}Sr_{0.55}MnO_{2.80\sim2.77}$) constitute heterostructures of five different states, which are demonstrated by direct crystal structure characterizations, electronic structure analyses, and transport measurements. It is observed that a positive gate voltage is preferential to inject oxygen vacancies into perovskite $SrCoO_3$ to form brownmillerite $SrCoO_{2.5}$, while a negative gate voltage tends to first extract oxygen vacancies from LSMO, both in experiment and theory. The sequence of various states in this reversible manipulation is mainly determined by the oxygen vacancies formation energy and Gibbs free energy difference of SCO and LSMO. Our results not only realize the reversible manipulation of the electronic phase transition of SCO and LSMO by an electrical means but also provide further understanding of the oxygen vacancies formation and annihilation in complex bilayer oxide systems.

## ACKNOWLEDGMENTS

We acknowledge Beamline BL08U1A in Shanghai Synchrotron Radiation Facility (SSRF) for XAS measurements. B.C. is grateful for the support of the Alexander von Humboldt Foundation. C.S. acknowledges the support of Young Chang Jiang Scholars Program, Beijing Advanced Innovation Center for Future Chip (ICFC). This work was supported by Ministry of Science and Technology of the People's Republic of China (Grant No. 2017YFB0405704).

## REFFERENCES

**Table 1 Phase composition and magnetic/electrical properties of different states controlled by electric-field.**

| State | Phase | *x* | *y* | *R* @ 300 K (Ω) | $T_C$ (K) | $H_C$ (kOe) | $H_{EB}$ (Oe) |
|---|---|---|---|---|---|---|---|
| ① | BM-SCO/L-LSMO | 0.49 | 0.23 | $6.5\times10^4$ | 283 | 0.15 | 50 |
| ② | BM-SCO/M-LSMO | 0.42 | 0.15 | $2.6\times10^4$ | 320 | 0.23 | 75 |
| ③ | P-SCO/M-LSMO | 0.15 | 0.12 | $6.8\times10^2$ | - | 2.00 | 0 |
| ④ | P-SCO/H-LSMO | 0.05 | 0.01 | $3.8\times10^3$ | - | 1.92 | 0 |
| ⑤ | BM-SCO/H-LSMO | 0.42 | 0.04 | $5.1\times10^5$ | 230 | 0.15 | 0 |
| ⑥ | BM-SCO/M-LSMO | 0.50 | 0.09 | $3.3\times10^4$ | 317 | 0.24 | 65 |
| ⑦ | BM-SCO/L-LSMO | 0.50 | 0.20 | $7.3\times10^4$ | 298 | 0.18 | 75 |

**Figure Captions**

**Fig. 1.** (a) Crystal structures of $SrCoO_{2.5}$ and $SrCoO_3$. (b) Electronic phase diagram of $La_{0.45}Sr_{0.55}MnO_{3-y}$ with varying $y$. The positions of L-, M-, and H-LSMO are marked by the circles in the phase diagram, which are determined by the results shown in Table 1 below. The arrows indicate the effect of gate voltage on the phase transition of SCO and LSMO. (c) The schematic cross-section view along the thin film device for IL gating ($V_G > 0$). (d) Schematic diagram of the transistor device with an ionic liquid electrolyte. The source, drain, and gate electrodes are denoted as S, D, and G, respectively.

**Fig. 2.** (a) X-ray diffraction patterns for heterostructures in different states. (b) and (c) STEM-HAADF of heterostructures (left) and -ABF of SCO layer (right) in states ① and ④ respectively, seen along the [110] STO direction. The higher the atomic number is, the brighter (darker) the atom in HAADF (ABF). The Co, Sr, O, and $V_O$ sites are indicated by the orange, green, red spheres and red circles, respectively.

**Fig. 3.** (a) The Co-$L_3$ (left axis) and Mn-$L_3$ (right axis) peak position of every state in XAS at 300 K. The states with positive and negative $V_G$ are highlighted by pink and soft blue shallow, respectively. The error fluctuation in XAS peak position is only ~6.7% according to the measurements of 5 times. (b) O-$K$ edge for all the states at 300 K. The curves are normalized according to the intensity of peak A and the area of peak B is highlighted by the yellow shallow.

**Fig. 4.** (a) Schematic for the $La_{0.5}Sr_{0.5}MnO_3/SrCoO_3$ superlattice. The oxygen ions are classified into 16 different cases according to the chemical situation and marked by the numbers. (b) The energy of the superlattice with oxygen vacancy at different positions. The interface between SCO and LSMO is marked by the dashed line.

Fig. 5. (a) Sketch for the reversible manipulation of electrical control of phase transition in SCO/LSMO heterostructures. (b) Channel resistance ($R$) versus $V_G$ at 300 K. The positions of seven states are marked by the numbers. The inset shows that the conduction path of SCO/LSMO heterostructure is the parallel circuit of SCO and LSMO layers. (c) Temperature dependent channel resistances ($R$–$T$) for the seven states. (d) The $T_C$ of every state extracted from the $R$–$T$ curves. (e) The $H_C$ (left axis) and $H_{EB}$ (right axis) of every state according to the transport measurement at 10 K (raw data in Fig. S11 of [29]).

**Fig. 6.** Sketch for the oxygen vacancy formation energies ($E_V$, red arrows/values), Gibbs free energy difference ($\Delta G$, dark cyan arrows/values), and the corresponding energy barriers for different phase transition ($E_B$, red and blue arrows/values for reduction and oxidation phase transitions, respectively) in SCO and LSMO. The oxygen vacancy formation energies are 1.70 eV for P-SCO and 2.70 eV for H-LSMO, which are the energy barriers for the reduction phase transition of P-SCO → BM-SCO and H-LSMO → M-LSMO [37,38], respectively. The energy barriers for the oxidation phase transition of BM-SCO → P-SCO (2.32 eV) and M-LSMO → H-LSMO (2.55 eV) are determined by the sum of oxygen vacancy formation energies and Gibbs free energy difference [8]. The absolute height for oxygen vacancy formation energy is the same in $La_{0.45}Sr_{0.55}MnO_{3-y}$, thus the energy barrier for M-LSMO → L-LSMO is 2.55 eV and L-LSMO → M-LSMO is 2.18 eV taking their Gibbs free energy difference into consideration.

**Fig. 7.** (a) The Co-$L_3$ (left axis) and Mn-$L_3$ (right axis) peak position of every state in XAS at 300 K. (b) Sketch for the reversible manipulation of electrical control of phase transition in LSMO/SCO heterostructure. The states with positive and negative $V_G$ are highlighted by pink and soft blue shallow, respectively.

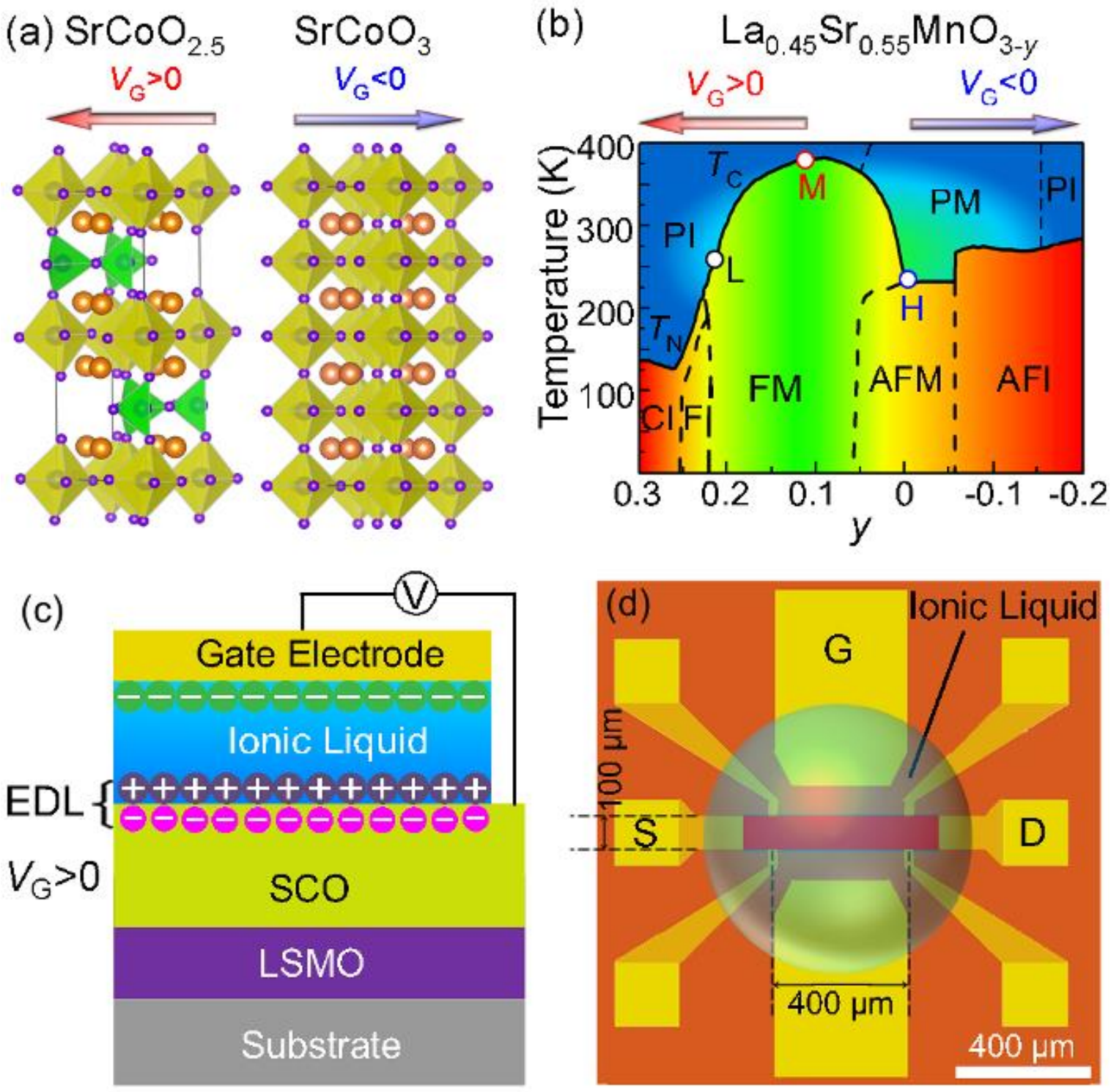

Cui *et al*., FIG. 1

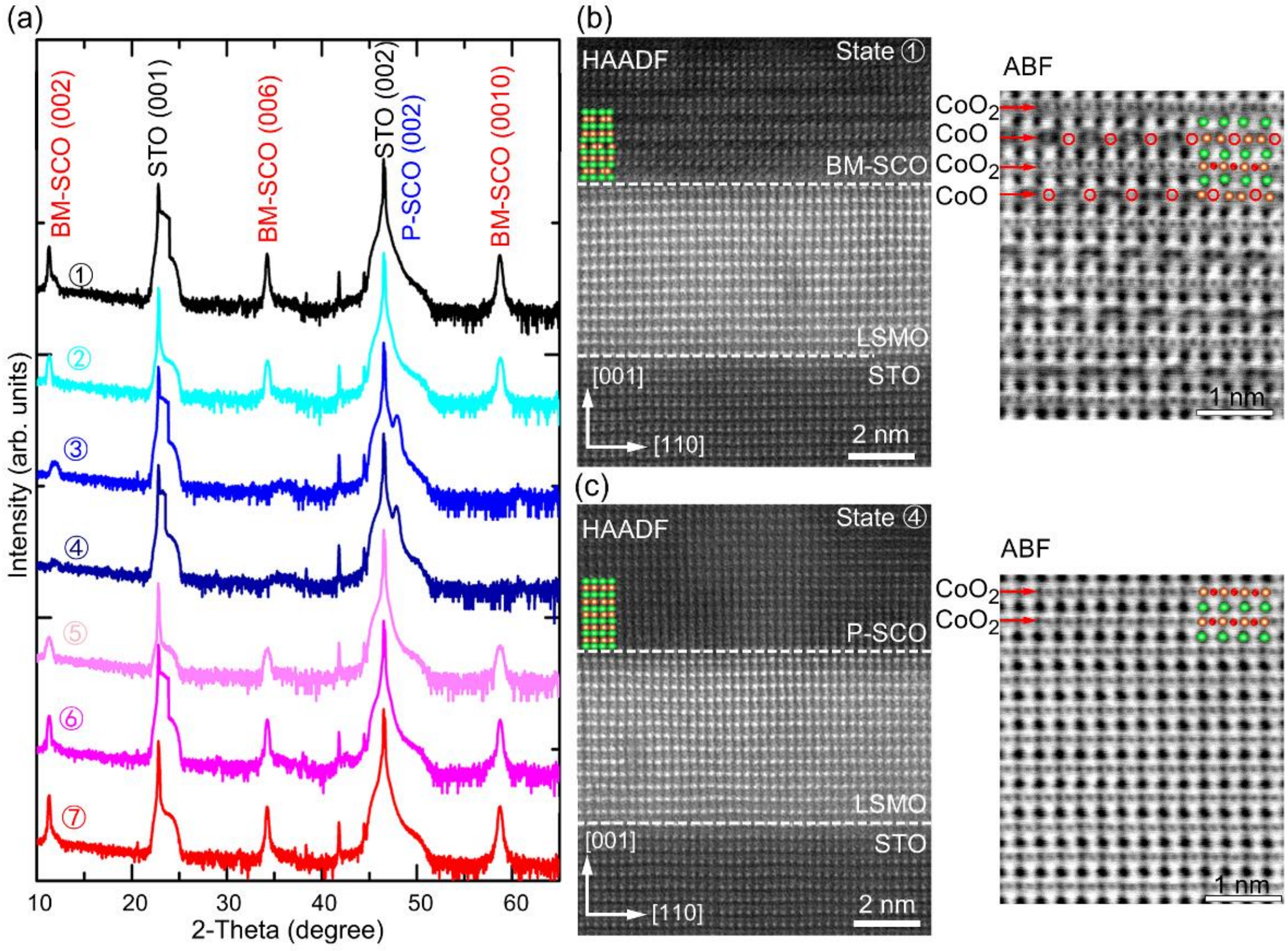

Cui *et al*., FIG. 2

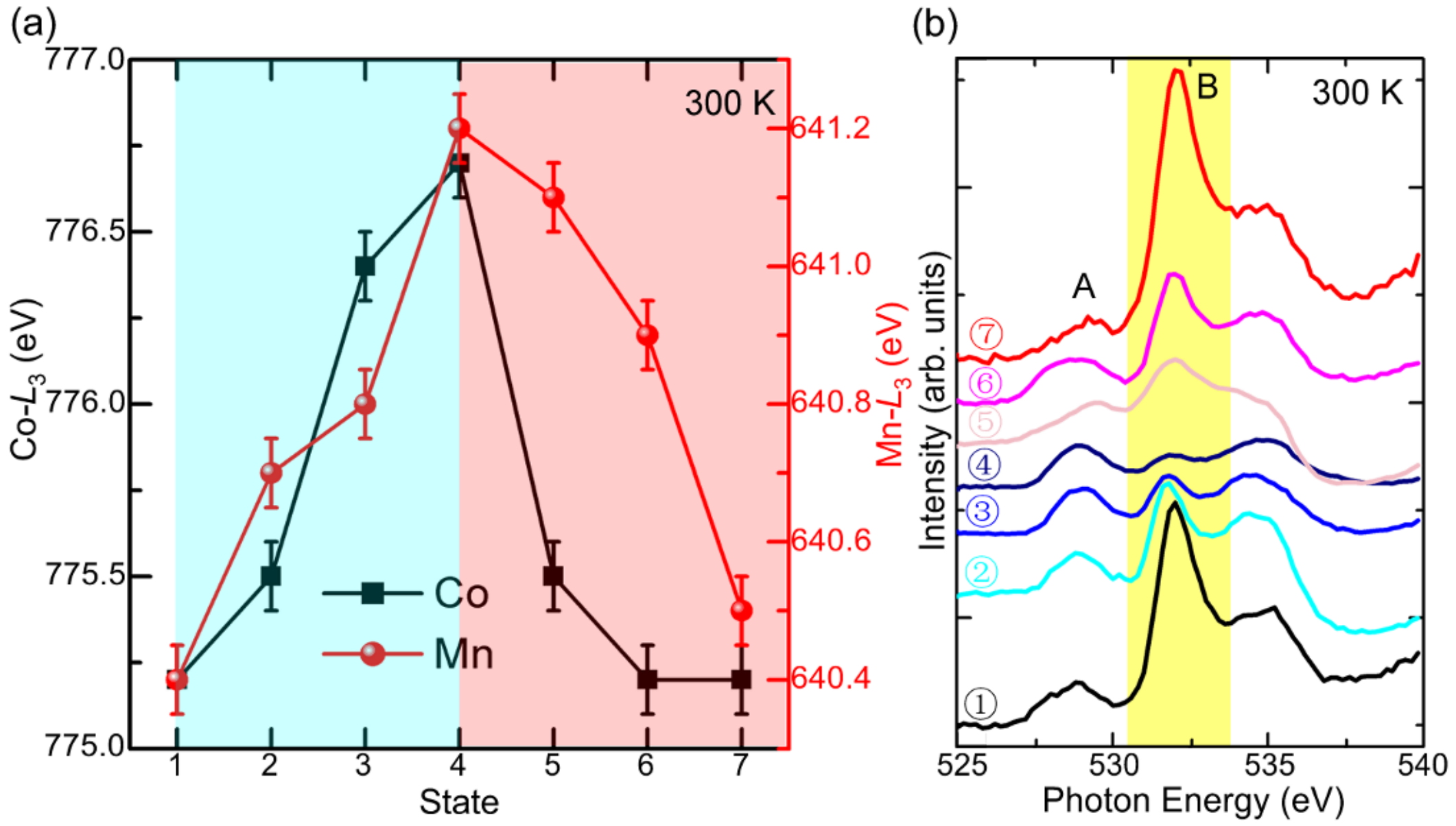

Cui *et al*., FIG. 3

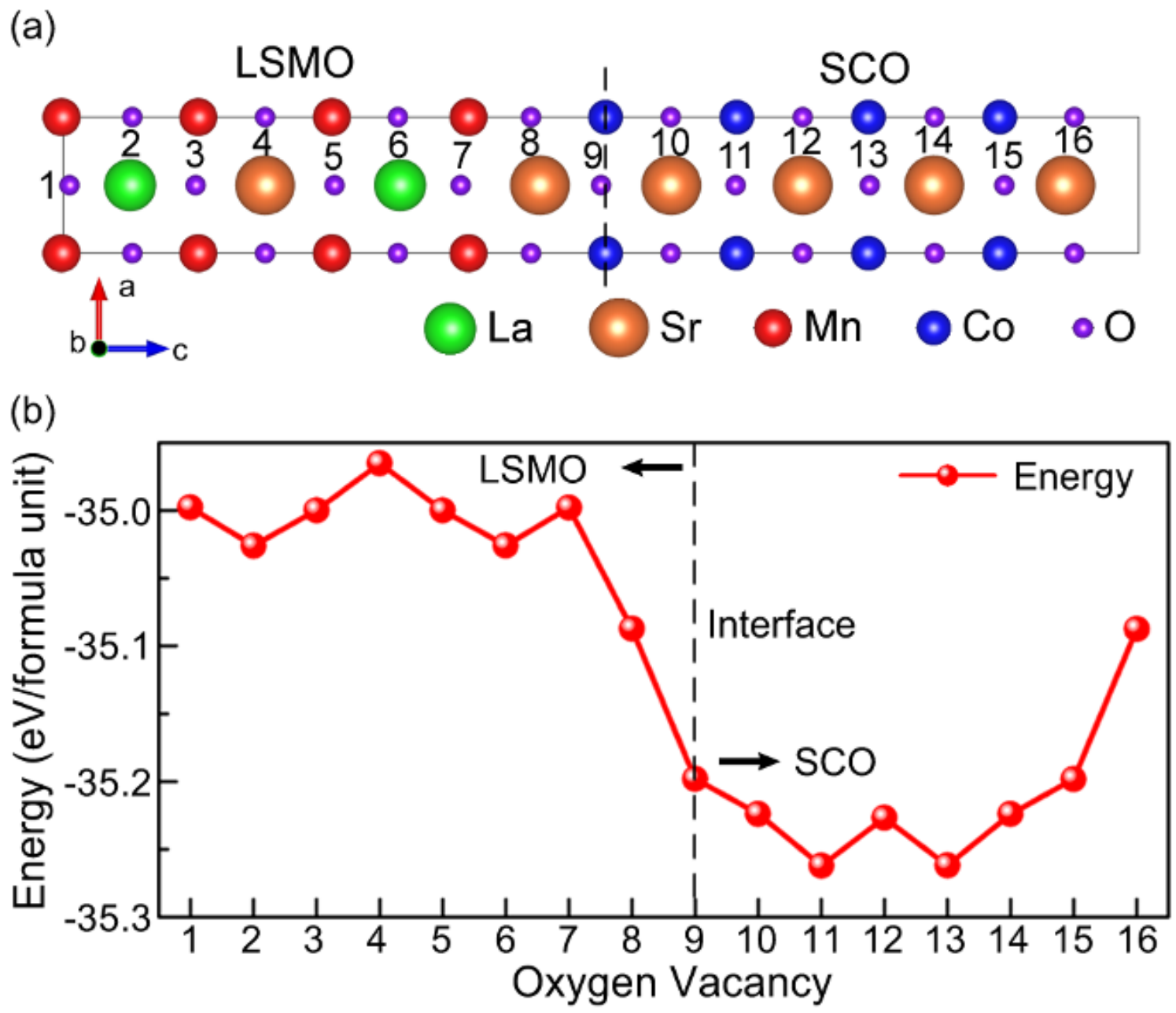

Cui *et al*., FIG. 4

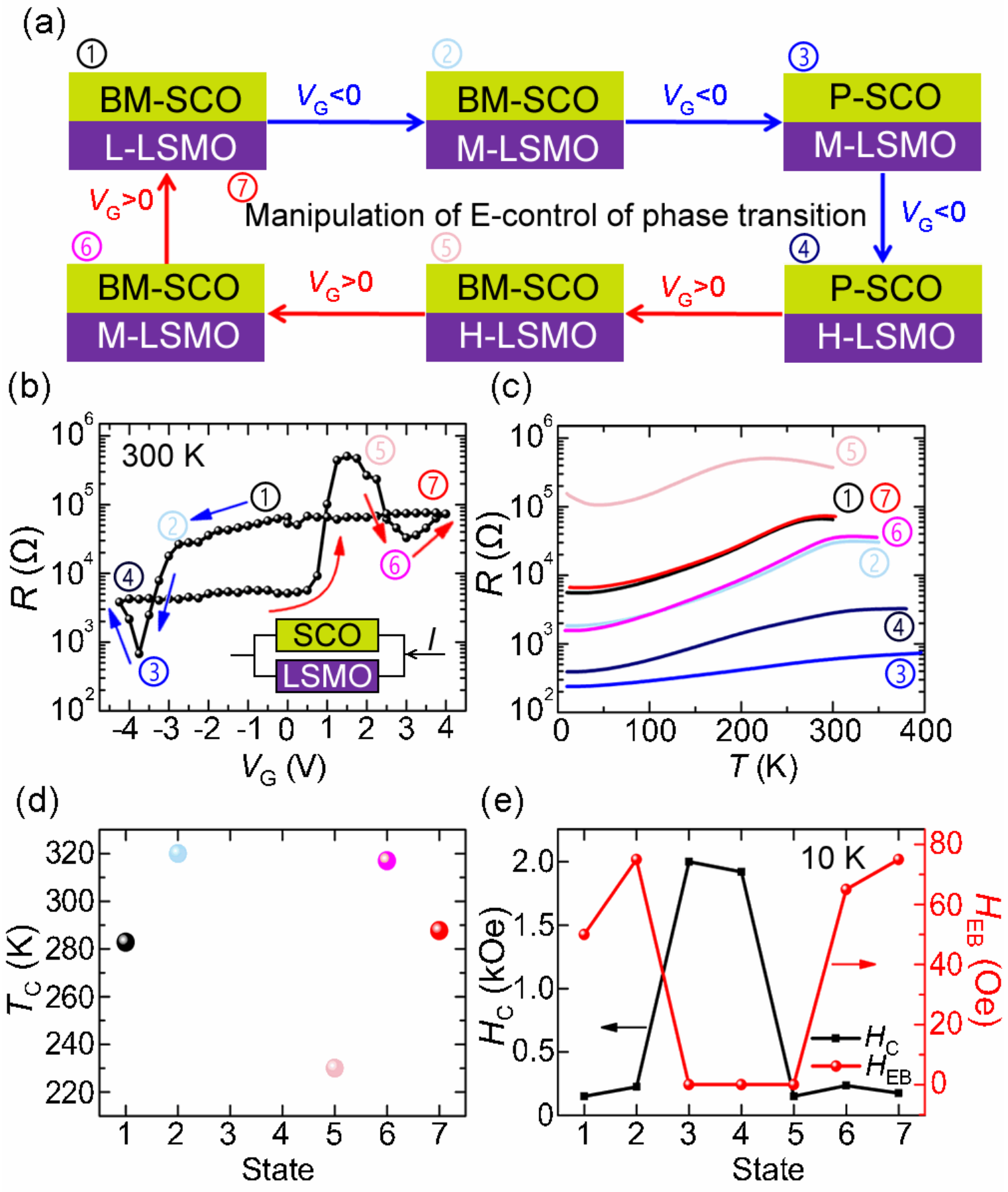

Cui *et al*., FIG. 5

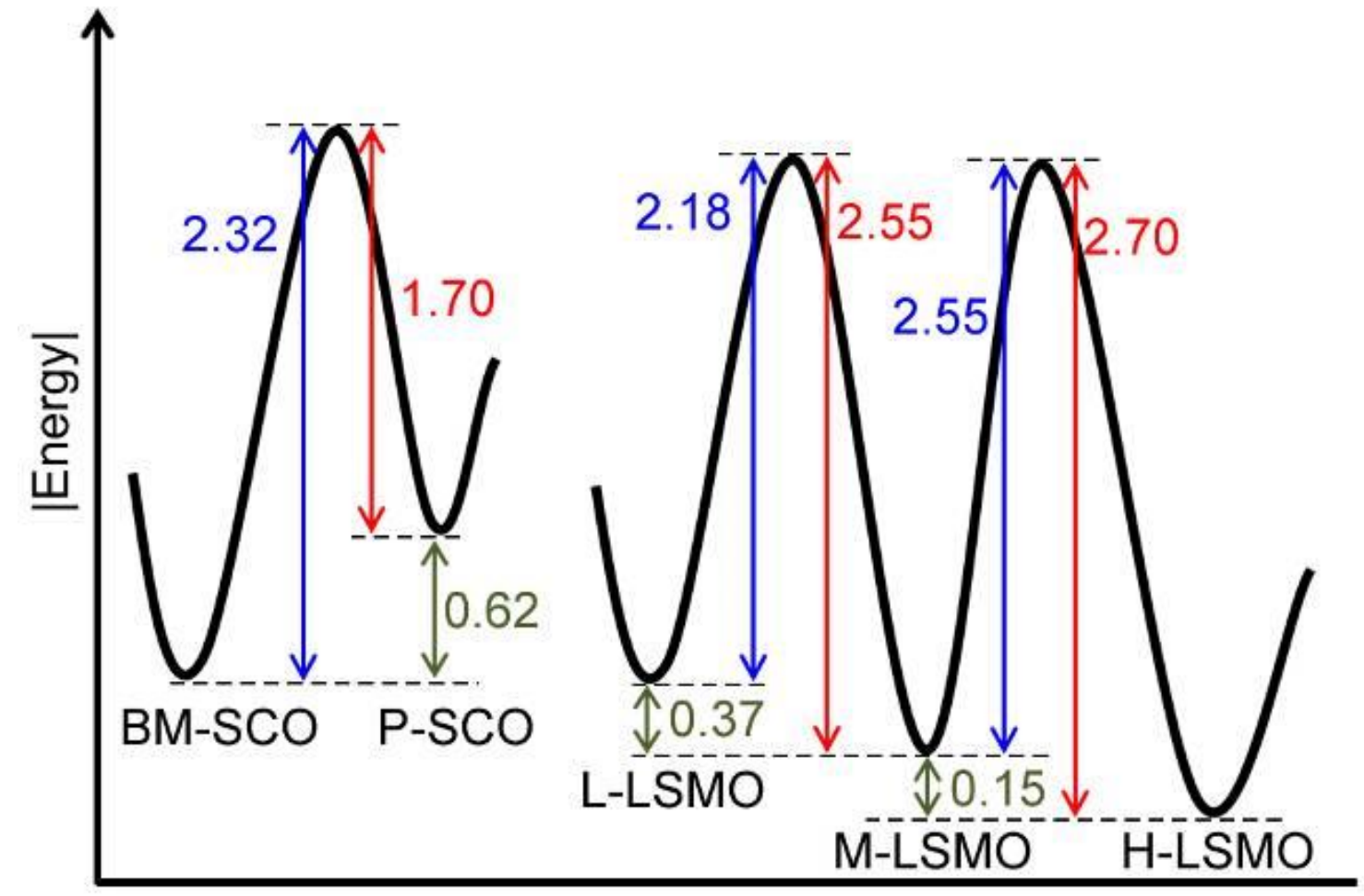

Cui *et al.*, FIG. 6

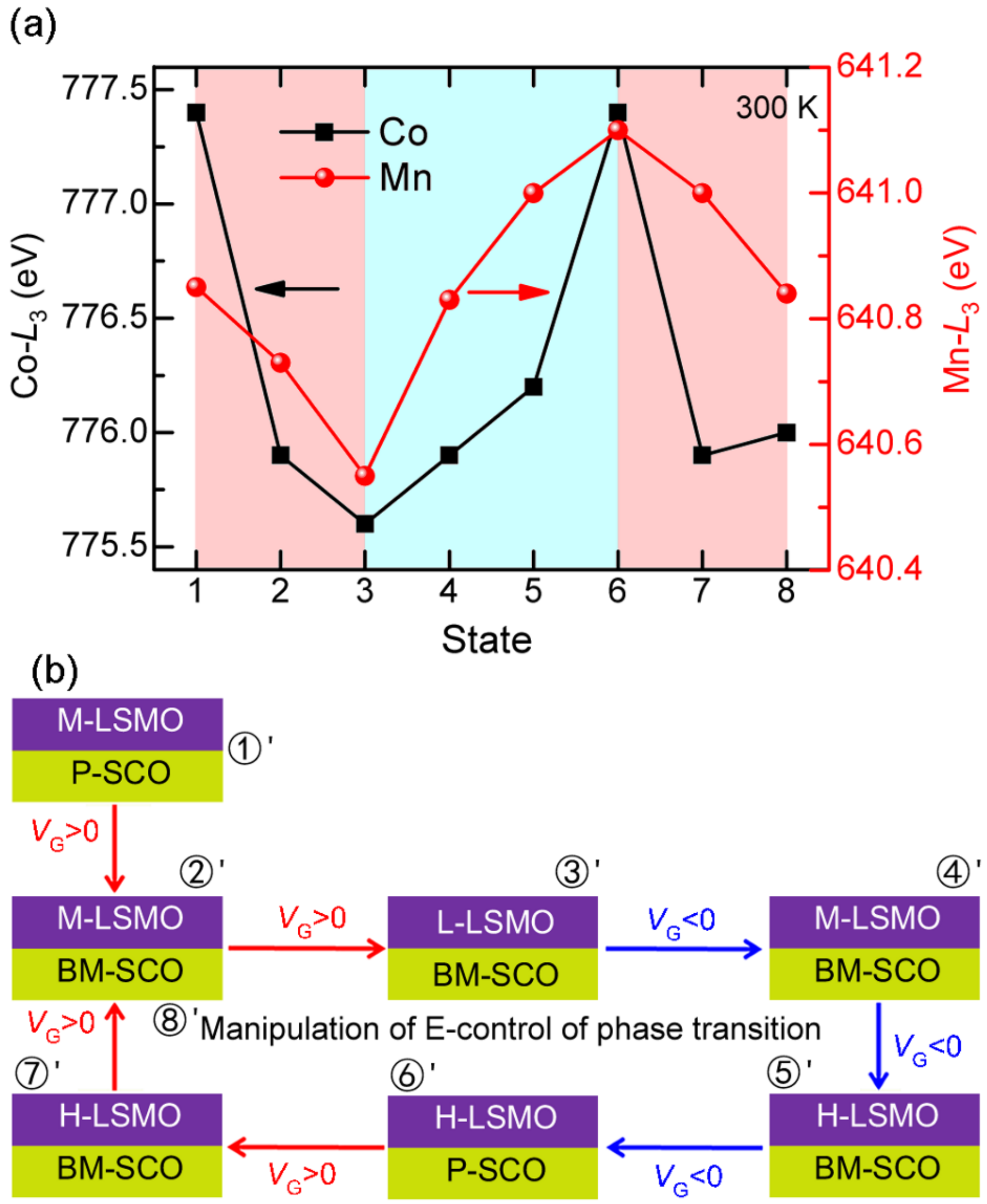

Cui *et al.*, FIG. 7